\documentclass[twocolumn,showpacs,preprintnumbers,amssymb,prl]{revtex4}    
\usepackage{graphicx}
\usepackage{color}
\usepackage{dcolumn}
\usepackage{bm}
\usepackage{longtable}
\usepackage{graphics}
\begin{document}

\title{Fine-grained uncertainty relation and nonlocality of tripartite systems}

\author{T. Pramanik\footnote{tanu.pram99@bose.res.in}}
\affiliation{S. N. Bose National Centre for Basic Sciences, Salt Lake, Kolkata 700 098, India}

\author{A. S. Majumdar\footnote{archan@bose.res.in}}
\affiliation{S. N. Bose National Centre for Basic Sciences, Salt Lake, Kolkata 700 098, India}

\begin{abstract}
The upper bound of the fine-grained uncertainty relation is  different for classical physics, quantum physics and no-signaling theories with maximal nonlocality (supper quantum correlation), as was shown  in the case of bipartite systems [J. Oppenheim and S. Wehner, Science {\bf 330}, 1072 (2010)]. Here, we extend the fine-grained uncertainty relation to the case of tripartite systems. We show that the fine-grained uncertainty relation determines the nonlocality of tripartite systems as manifested by the Svetlichny inequality, discriminating between classical physics, quantum physics and super quantum correlations.
\end{abstract}

\pacs{03.65.Ud, 03.65.Ta}

\maketitle

Uncertainty relations prohibit our complete knowledge about the physical properties described by the state of a system. According to the Heisenberg uncertainty relation \cite{HUR},  we are unable to predict certainly the measurement outcomes of two non-commutating observables. For example, when one predicts certainly the spin orientation of a qubit along the z-axis, the knowledge of spin orientation of that qubit along the x-axis (or, the y-axis) is completely uncertain, i.e., the probability of getting spin up and down are equal (to 1/2). Schr$\ddot{o}$dinger and Robertson \cite{GUR} generalized the uncertainty relation for any two arbitrary observables. Subsequently, a number of works have been done in the direction of separating entangled states from separable states using several forms of the uncertainty relation \cite{Nha,Ag1,SWYP}. 

In quantum information theory, an uncertainty relation in terms of entropy is more useful than that in terms of standard deviation. Much effort has been devoted towards improvement of the entropic uncertainty relation \cite{Sur,IBB,Deu}. Entropic uncertainty relations are used to find out the security key rate in cryptographic protocols, and recently, Berta et al. \cite{Berta} have introduced a new lower bound of entropic uncertainty depending upon the amount of entanglement between the particle and quantum memory. This is applicable in cryptographic scenarios \cite{cryp}, locking information \cite{lock} and decoupling theorems that are used in coding arguments \cite{code}.

In a recent work, Oppenheim and Wehner \cite{Oppenheim} have proposed a new fine-grained uncertainty relation which is aimed at capturing the plurality of
simultaneous possible outcomes of a set of measurements. Considering bipartite
systems they have exemplified such an uncertainty relation for a special class
of nonlocal retreival games for which there exist only one winning answer for
one of the two parties. The upper bound of the uncertainty relation which is
also the maximum winning probability of the retrieval game was shown to
have an implication on the degree of nonlocality of the underlying physical
theory. In particular, such an upper bound could be used to discriminate
between the degree of nonlocality pertaining to classical theory, quantum
theory, and no-signalling theory with maximum 
nonlocality for bipartite systems. 

In the present paper we
investigate a possible connection between fine-grained uncertainty and 
nonlocality for tripartite systems. Unlike the bipartite case where 
correlations are unambiguously expressible in the Bell-CHSH form 
\cite{bell, bell2}, in tripartite systems there is an 
inherent non-uniqueness regarding 
the choice of the type of correlations proposed by Svetlichny \cite{Svet} and 
Mermin \cite{Mermin}.  Here we study this issue by generalizing the 
fine-grained uncertainty relation for the case of tripartite systems.

Let us first consider the fine-grained uncertainty relation as proposed
by Oppenheim and Wehner \cite{Oppenheim}
of the form given by
\begin{eqnarray}
P(\sigma ,\textbf{x}):= \displaystyle\sum_{t=1}^n p(t) p(x^{(t)}|t)_{\sigma} \leq \zeta_{\textbf{x}}(\mathcal{T},\mathcal{D})
\label{FUR1}
\end{eqnarray}
where $P(\sigma ,\textbf{x})$ is the total probability of possible outcomes written as a string $\textbf{x}=\{x^{(1)}, ..., x^{(n)}\}$ corresponding to a set of measurements $\{t\}$ $(\in \mathcal{T})$ chosen with probabilities $\{p(t)\}$ ($\in \mathcal{D}$, the probability distribution of choosing measurements), $p(x^{(t)}|t)_{\sigma}$ is the probability of obtaining outcome $x^{(t)}$ by performing measurement labeled `t' on the state of a general physical system $\sigma$, $n (=|\mathcal{T}|)$ is the total number of different measurement  settings, and $\zeta_{\textbf{x}}(\mathcal{T},\mathcal{D})$ is given by
\begin{eqnarray}
\zeta_{\textbf{x}}(\mathcal{T},\mathcal{D})= \max_{\sigma} \displaystyle\sum_{t=1}^n p(t) p(x^{(t)}|t)_{\sigma}
\label{maxfur}
\end{eqnarray}
where the maximum is taken over all possible states allowed on a particular system. The uncertainty of measurement outcome occurs for the value of $\zeta_{\textbf{x}}(\mathcal{T},\mathcal{D})<1$. The value of $\zeta_{\textbf{x}}(\mathcal{T},\mathcal{D})$ is bound by the particular physical theory. For example, no-signaling theory with maximum nonlocality gives the upper bound $\zeta_{\textbf{x}}(\mathcal{T},\mathcal{D})=1$ \cite{Oppenheim}.

It is instructive to consider specific forms of the above uncertainty relation 
that correspond to specific choices of systems.  For example, for the case of
the single qubit in quantum theory, the form of the fine-grained uncertainty 
relation is given by \cite{Oppenheim}
\begin{eqnarray}
P(\mathcal{T},\sigma_A)=\displaystyle\sum_{t=1}^n p(t) p(a=x^{(t)}|t)_{\sigma_A}\leq \zeta_{\textbf{x}}(\mathcal{T},\mathcal{D})
\end{eqnarray}
where $p(a=x^{(t)}|t)_{\sigma_A}$ is given by
\begin{eqnarray}
p(a=x^{(t)}|t)_{\sigma_A}=Tr[A_t^a.\sigma_A]
\end{eqnarray}
with $A_t^a$  being the measurement operator corresponding to measurement setting `t' giving outcome `a', and $\zeta_{\textbf{x}}(\mathcal{T},\mathcal{D})$ is given by
\begin{eqnarray}
\zeta_{\textbf{x}}(\mathcal{T},\mathcal{D})=\max_{\sigma_A} P(\mathcal{T},\sigma_A).
\end{eqnarray}
Here the maximum is taken over all possible single qubit states.
The value of $\zeta_{\textbf{x}}(\mathcal{T},\mathcal{D})$ that occurs for the spin measurements along the z-axis (by measuring the observable $\sigma_z$) and along the x-axis (by measuring the  observable $\sigma_x$) with equal probability (i.e., p(t)=1/2) on the eigenstates of $(\sigma_x+\sigma_z)/\sqrt{2}$ and $(\sigma_x-\sigma_z)/\sqrt{2}$, is $(\frac{1}{2} + \frac{1}{2 \sqrt{2}})$.
 
An interesting connection between the fine-grained uncertainty relation and
nonlocality was observed by Oppenheim and Wehner \cite{Oppenheim} for the
case of bipartite systems. They provided specific examples of 
nonlocal retrieval games (for which there exist only one winning answer for
one of the two parties) for the purpose of discriminating different types
of theories by the upper bound of $\zeta$ (the degree of nonlocality). 
According to these games, Alice and Bob receive questions `s' and `t' respectively, with some probability distribution $p(s,t)$ (for simplicity, $p(s,t)=p(s) p(t)$); and their answer `a' or `b' will be winning answers determined by the 
set of rules, i.e., for every setting `s' and the corresponding outcome `a' of 
Alice, there is a string $\textbf{x}_{s,a}=(x_{s,a}^{(1)}, ..., x_{s,a}^{(n)})$ of 
length $n=|\mathcal{T}|$ that determines the correct answer $b=x_{s,a}^{t}$ for 
the question `t' for Bob. In the prescribed game (CHSH game), Alice and Bob 
receive respective binary questions $s,t \in \{0,1\}$ (i.e., representing two different measurement settings on each side), and they win the game if their 
respective outcomes (binary) $a,b\in \{0,1\}$ satisfy the condition $a\oplus b=s.t$. At the starting of the game, Alice and Bob discuss their strategy (i.e., choice of shared bipartite state and also measurement). They are not allowed to
communicate with each other once the game has started. The probability of 
winning the game for a physical theory described by bipartite state ($\sigma_{AB}$) is given by
\begin{eqnarray}
P^{game}(\mathcal{S},\mathcal{T},\sigma_{AB})=\displaystyle\sum_{s,t} p(s,t) \displaystyle\sum_a p(a,b=x_{s,a}^t|s,t)_{\sigma_{AB}}
\label{FUR2}
\end{eqnarray}
where the form of $p(a,b=x_{s,a}^t|s,t)_{\sigma_{AB}}$ in terms of the measurements on the bipartite state $\sigma_{AB}$ is given by
\begin{eqnarray}
p(a,b=x_{s,a}^t|s,t)_{\sigma_{AB}}= \displaystyle\sum_b V(a,b|s,t) \langle (A_s^a\otimes B_t^b)\rangle_{\sigma_{AB}}
\label{prob2}
\end{eqnarray}
where $A_s^a$ ($=\frac{(\mathcal{I}+(-1)^a A_s)}{2}$) is a measurement of the
observable $A_s$ corresponding to setting `s' giving outcome `a' at Alice's side; $B_t^b$ ($=\frac{(\mathcal{I}+(-1)^a B_s)}{2}$) is a measurement of the
observable $B_t$ corresponding to setting `t'  giving outcome `b' at Bob's 
side, and $V(a,b|s,t)$ is the winning condition given by
\begin{eqnarray}
V(a,b|s,t)&=& 1 \text{\phantom{xxxxx} iff $a\oplus b = s.t$} \nonumber \\
          &=& 0 \text{\phantom{xxxxx} otherwise}
\label{cond2}
\end{eqnarray}
Using  Eqs. (\ref{FUR2}), (\ref{prob2}), (\ref{cond2}) and taking $p(s,t)=p(s)p(t)=1/4$,  one can get the expression of $P^{game}(\mathcal{S},\mathcal{T},\sigma_{AB})$ for the bipartite state, $\sigma_{AB}$ given by
\begin{eqnarray}
P^{game}(\mathcal{S},\mathcal{T},\sigma_{AB})=\frac{1}{2}(1+\frac{\langle\mathcal{B}_{CHSH}\rangle_{\sigma_{AB}}}{4})
\end{eqnarray}
where $\mathcal{B}_{CHSH}=A_0\otimes B_0+A_0\otimes B_1+A_1\otimes B_0-A_1\otimes B_1$, and corresponds to the well-known Bell-CHSH operator \cite{bell,bell2}.
To characterize the allowed distribution under the theory, we need to know the maximum winning probability, maximized over all possible strategies for Alice and Bob. The maximum winning probability is given by
\begin{eqnarray}
P^{game}_{\max} = \max_{\mathcal{S},\mathcal{T},\sigma_{AB}} P^{game}(\mathcal{S},\mathcal{T},\sigma_{AB})
\label{maxgame}
\end{eqnarray}
The value of $P^{game}_{\max}(\mathcal{S},\mathcal{T},\sigma_{AB})$ allowed by classical physics is $\frac{3}{4}$ (as classically, the Bell-CHSH inequality  is bounded by $2$), by quantum mechanics is $(\frac{1}{2}+\frac{1}{2 \sqrt{2}})$ (due to the maximum violation of Bell inequality,  $\langle \mathcal{B}_{CHSH} \rangle =2\sqrt{2}$), and by no-signaling theories with maximum Bell violation ($\langle \mathcal{B}_{CHSH}\rangle=4$, that occurs for the PR-box \cite{PR}) is $1$. The connection of
Eq.(\ref{cond2}) with the no-signalling constraint for the general case of a 
bipartite system was elaborated by Barrett et al. \cite{nosignal}. The
connection between the bound on the fine-grained uncertainty relation and the 
maximum degree of nonlocality in a given physical theory is thus established 
by the correspondence
of Eq.(\ref{maxfur}) with Eq.(\ref{maxgame})

We now generalize the fine-grained uncertainty relation 
for the case of tripartite systems,
classifying different no-signaling theories on the basis of their degree  of 
nonlocality. For tripartite systems we consider a nonlocal retrieval game 
similar to CHSH-game for bipartite systems.  Here, Alice, Bob and Charlie 
receive respective binary questions `s', `t', and  `u' $\in\{0,1\}$ (corresponding to their two different measurement settings at each side), and they win the 
game if their respective outcomes (binary) `a', `b', and `c' $\in\{0,1\}$ satisfy certain rules. We limit our analysis to the three kinds of no-signaling 
boxes, known as full-correlation boxes, for which all one-party and two-party  
correlation in the system vanishes \cite{nosignal}. The
 game is won if their answers satisfy the set of rules,
either 
\begin{eqnarray}
a\oplus b \oplus c=s.t \oplus t.u \oplus u.s
\label{box1}
\end{eqnarray}
or 
\begin{eqnarray}
a\oplus b \oplus c=s.t \oplus s.u
\label{box2}
\end{eqnarray}
or else
\begin{eqnarray}
a\oplus b \oplus c=s.t.u
\label{box3}
\end{eqnarray}
All the above boxes violate the Mermin inequality \cite{Mermin}, whereas the 
Svetlichny inequality \cite{Svet} is violated only by the box given by Eq. (\ref{box1}) (known as the Svetlichny box). The winning probability of our prescribed game under a physical theory described by a shared tripartite state $\sigma_{ABC}$ (among Alice, Bob and Charlie) is given by
\begin{eqnarray}
&& P^{game}(\mathcal{S},\mathcal{T},\mathcal{U},\sigma_{ABC})\nonumber \\
&&=\displaystyle\sum_{s,t,u} p(s,t,u) \displaystyle\sum_{a,b} p(a,b,c=x_{s,t,a,b}^{(u)}|s,t,u)_{\sigma_{ABC}}
\label{FUR3}
\end{eqnarray}
where $p(s,t,u)$ is the probability of choosing the measurement settings `s' by Alice, `t' by Bob and `u' by Charlie, and  $p(a,b,c|s,t,u)_{\sigma_{ABC}}$ the joint probability of getting outcomes `a', `b' and `c' for corresponding settings `s', `t' and `u' given by
\begin{eqnarray}
&& p(a,b,c=x_{s,t,a,b}^{(u)}|s,t,u)_{\sigma_{ABC}}  \nonumber \\
&&  =\displaystyle\sum_{c}  V(a,b,c|s,t,u) \langle A_s^a\otimes B_t^b\otimes C_u^c\rangle_{\sigma_{ABC}}
\label{prob3}
\end{eqnarray}
where $A_s^a$, $B_t^b$ and $C_u^c$ are the measurements corresponding to setting `s' and outcome `a' at Alice's side, setting `t' and outcome `b' at Bob's side, and setting `u' and outcome `c' at Charlie's side, respectively; and $V(a,b,c|s,t,u)$ (the winning condition) is non zero ($=1$) only when the outcomes of Alice, Bob and Charlie are correlated by either of Eqs. (\ref{box1}), (\ref{box1}) or (\ref{box3}), and is zero otherwise. The maximum winning probability over all 
possible strategies (i.e., the choice of the shared tripartite state and 
measurement settings by the three parties) for any theory is given by
\begin{eqnarray}
P^{game}_{\max} = \max_{\mathcal{S},\mathcal{T},\mathcal{U},\sigma_{ABC}} P^{game}(\mathcal{S},\mathcal{T},\mathcal{U},\sigma_{ABC})
\end{eqnarray}
which is a signature of the allowed probability distribution under that theory.
In the following we study the cases of classical, qauntum and no-signalling
theories with super-quantum correlations for the above different 
full-correlation boxes (rules of the nonlocal 
game) separately.  

For the case of the winning condition given by Eq. (\ref{box1}), and assuming 
$p(s,t,u)=p(s)p(t)p(u)=\frac{1}{8}$, the expression of $P^{game}(\mathcal{S},\mathcal{T},\mathcal{U},\sigma_{ABC})$ for the shared tripartite state $\sigma_{ABC}$ (among Alice, Bob and Charlie)  is given by
\begin{eqnarray}
P^{game}(\mathcal{S},\mathcal{T},\mathcal{U},\sigma_{ABC})=\frac{1}{2} [1+\frac{\langle \textbf{S}_1\rangle_{\sigma_{ABC}}}{8}]
\end{eqnarray}
where
\begin{eqnarray}
\textbf{S}_1=&&A_0\otimes B_0\otimes C_0+A_0\otimes B_0\otimes C_1+A_0\otimes B_1\otimes C_0 \nonumber \\
&& +A_1\otimes B_0\otimes C_0-A_0\otimes B_1\otimes C_1-A_1\otimes B_0\otimes C_1\nonumber \\
&&-A_1\otimes B_1\otimes C_0-A_1\otimes B_1\otimes C_1
\end{eqnarray}
The value of $P^{game}_{\max}$ allowed in classical physics is $3/4$ which 
follows from the Svetlichny inequality \cite{Svet}
\begin{eqnarray}
\langle \textbf{S}_1\rangle_{\sigma_{ABC}} \leq 4
\end{eqnarray}
For the case of quantum physics, we consider the two classes of pure entangled 
states for tripartite systems,  i.e., the GHZ state ($\frac{|000\rangle_{ABC} + |111\rangle_{ABC}}{\sqrt{2}}$) and the W state  ($\frac{|001\rangle_{ABC} +  |010\rangle_{ABC} + |100\rangle_{ABC}}{\sqrt{3}}$). The 
maximum violation of the Svetlichny inequality is $4\sqrt{2}$ which occurs for
the GHZ-state  \cite{S1GHZ}, whereas the violation of the Svetlichny inequality by the W-state is given by 4.354 \cite{S1W}. Hence, the value of $P^{game}_{\max}$ allowed in quantum physics is $(\frac{1}{2}+\frac{1}{2 \sqrt{2}})$. For the 
case of the maximum no-signalling theory, the algebraic maximum of the  Svetlichny inequality  is $8$ \cite{nosignal}, and the value of $P^{game}_{\max}$ in
this case
is $1$, corresponding to a correlation with maximum nonlocality.

Next, let us consider the winning condition  for the outcomes related by Eq. (\ref{box2}). In this case the expression of $P^{game}(\mathcal{S},\mathcal{T},\mathcal{U},\sigma_{ABC})$ is given by
\begin{eqnarray}
P^{game}(\mathcal{S},\mathcal{T},\mathcal{U},\sigma_{ABC})=\frac{1}{2} [1+\frac{\langle\textbf{S}_2\rangle_{\sigma_{ABC}}}{8}]
\end{eqnarray}
where
\begin{eqnarray}
\textbf{S}_2=&&A_0\otimes B_0\otimes C_0+A_0\otimes B_0\otimes C_1+A_0\otimes B_1\otimes C_0 \nonumber \\
&&+A_1\otimes B_0\otimes C_0+A_0\otimes B_1\otimes C_1-A_1\otimes B_0\otimes C_1\nonumber \\
&&-A_1\otimes B_1\otimes C_0+A_1\otimes B_1\otimes C_1
\end{eqnarray}
It can be seen that when all the variables $A_i$, $B_i$ and $C_i$ take either `+1' or `-1', the maximum value of $\langle\textbf{S}_2\rangle_{\sigma_{ABC}}$ is $4$, and hence, the value of $P^{game}_{\max}$ for classical physics is $\frac{3}{4}$. To find out the maximum value of $\langle\textbf{S}_2\rangle_{\sigma_{ABC}}$ in quantum physics, we maximize the $\langle\textbf{S}_2\rangle_{\sigma_{ABC}}=Tr[\textbf{S}_2.\sigma_{ABC}]$ over all possible projective spin measurements on both the
GHZ-state and the W-state. The two observables of which each party (Alice, Bob and Charlie) performs one measurement,  are of form
\begin{eqnarray}
\Pi_{\alpha_0} &=& \sin(\theta \alpha 0) \cos(\phi \alpha 0) \sigma_x + \sin(\theta \alpha 0) \sin(\phi \alpha 0) \sigma_y  \nonumber \\
&& + \cos(\theta \alpha 0) \sigma_z \nonumber \\
\Pi_{\alpha_1} &=& \sin(\theta \alpha 1) \cos(\phi \alpha 1) \sigma_x + \sin(\theta \alpha 1) \sin(\phi \alpha 1) \sigma_y \nonumber \\
&& + \cos(\theta \alpha 1) \sigma_z 
\end{eqnarray}
where $\Pi\alpha_0$ is the spin observable in direction $\{\theta \alpha 0, \phi \alpha 0\}$ and $\Pi\alpha_1$ is the spin observable in the direction $\{\theta \alpha 1, \phi \alpha 1\}$ for party $\alpha$ ($\in\{A,B,C\}$); and $\sigma_i$'s are Pauli spin matrices. It is found numerically (using Mathematica)  that the maximum value of  $\langle\textbf{S}_2\rangle_{\sigma_{ABC}}$ for the GHZ-state as
well as the W-state is $4$. For example, the $P^{game}_{\max}$ occurs for the GHZ-state for the projective measurements either along the direction $\{\theta_{A0}=3.1149,\phi_{A0}=2.5271\}$ or along $\{\theta_{A1}=1.5708,\phi_{A1}=0.4608\}$ by Alice; along the direction
$\{\theta_{B0}=1.5708,\phi_{B0}=1.7282\}$ or along $\{\theta_{B1}=4.7124,\phi_{B1}=1.7282\}$ by Bob; and along the direction
$\{\theta_{C0}=4.7124,\phi_{C0}=0.9526\}$ or along $\{\theta_{C1}=4.7124,\phi_{C1}=4.0942\}$ by Charlie.
The value of  $P^{game}_{\max}$ turns out to be $\frac{3}{4}$ under the winning condition given by Eq. (\ref{box2}). Hence, this particular full corelation Mermin 
box \cite{Mermin} is
unable to distinguish classical theory from quantum theory in terms of their
degree of nonlocality. Nonetheless, similar to the case of the Svetlichny box
\cite{Svet} given by Eq.(\ref{box1}), the value of $P^{game}_{\max}$ in 
nosignaling theory with maximum nonlocality is $1$ here too.

Finally, we consider the winning condition given by the other Mermin box 
(\ref{box3}) and obtain
 the value of $P^{game}_{max}$ under different physical theories. The expression 
for $P^{game}(\mathcal{S},\mathcal{T},\mathcal{U},\sigma_{ABC})$ is given by
\begin{eqnarray}
P^{game}(\mathcal{S},\mathcal{T},\mathcal{U},\sigma_{ABC})=\frac{1}{2} [1+\frac{\langle\textbf{S}_3\rangle_{\sigma_{ABC}}}{8}]
\end{eqnarray}
where
\begin{eqnarray}
\textbf{S}_3=&&A_0\otimes B_0\otimes C_0+A_0\otimes B_0\otimes C_1+A_0\otimes B_1\otimes C_0 \nonumber \\
&&+A_1\otimes B_0\otimes C_0+A_0\otimes B_1\otimes C_1+A_1\otimes B_0\otimes C_1\nonumber \\
&&+A_1\otimes B_1\otimes C_0-A_1\otimes B_1\otimes C_1
\end{eqnarray}
In this case the maximum value of $\langle \textbf{S}\rangle_{\sigma_{ABC}}$ is $6$ for classical theory, and hence, the value of $P^{game}_{max}$ is $\frac{7}{8}$. In quantum mechanics, the maximum value of $\langle\textbf{S}\rangle_{\sigma_{ABC}}$ is determined by maximization over all possible 
projective measurements numerically, as discussed above. It turns out that for 
both the GHZ-state and the W-state the maximum value turns out to be $6$. 
The value of $P^{game}_{max}$ in quantum mechanics is thus $\frac{7}{8}$. Here,
 $P^{game}_{\max}$ occurs for the GHZ-state corresponding to  the projective 
measurements either along the direction 
$\{\theta_{A0}=4.7124,\phi_{A0}=1.6707\}$ or along $\{\theta_{A1}=4.7124,\phi_{A1}=1.6737\}$ by Alice; 
$\{\theta_{B0}=1.5708,\phi_{B0}=4.6120\}$ or along $\{\theta_{B1}=4.7124,\phi_{B1}=1.4735\}$ by Bob; and
$\{\theta_{C0}=4.7124,\phi_{C0}=6.2806\}$ or along $\{\theta_{C1}=4.7124,\phi_{C1}=4.0005\}$ by Charlie. Thus, the full-correlation Mermin box (\ref{box3}) also
fails to distinguish between classical and quantum physics using the
fine-grained uncertainty relation. However, one can see again that
in nosignaling theory with maximum nonlocality, the maximum value of $\langle\textbf{S}_3\rangle_{\sigma_{ABC}}$ is $8$ \cite{nosignal} corresponding to the value 
 $1$ for $P^{game}_{\max}$.

To summarize, in the present work we have generalized the connection between the fine-grained uncertainty relation, as expressed in terms of the maximum 
winning probability of prescribed retrieval 
nonlocal games,  and the degree of nonlocality of the underlying physical theory
\cite{Oppenheim} to the case of tripartite systems. We have shown that the 
fine-grained uncertainty relation determines the degree of nonlocality as manifested by the Svetlichny inequality for tripartite systems corresponding to the wining condition given by Eq. (\ref{box1}), in the same way as it determines the nonlocality of bipartite systems manifested by Bell-CHSH inequality. Thus, 
with the 
help of the fine-grained uncertainty relation, one is able to differentiate the 
various classes of theories (i.e., classical physics, quantum physics and no-signaling theories with maximum nonlocality or superquantum correlations) by the value of $P^{game}_{\max}$  for tripartite systems.  Further, within quantum theory
it is clear from the upper bound of $P^{game}_{\max}$ that the GHZ-state is more 
nonlocal than the W-state 
\cite{S1GHZ, S1W}. Finally, it may be noted that by 
considering the winning conditions of the
tripartite games given by the other two full correlation boxes, i.e., Eq. (\ref{box2}) and (\ref{box3}), which  violate the Mermin inequality but not the Svetlichny inequality,
the fine-grained uncertainty relation is unable to discriminate quantum 
physics from classical physics in terms of the degree of nonlocality.

\vskip 0.1in

{\it  Acknowledgements:}  The authors would like to thank Guruprasad Kar for helpful discussions.  TP thanks UGC, India for financial support and ASM acknowledges support from the DST project no. SR/S2/PU-16/2007.

\pagebreak

\end{document}